\documentstyle[prb,aps,epsf,twocolumn]{revtex}

\begin{document}

\wideabs{%
\title{On Quantum Effects near the Liquid--Vapor Transition in Helium}

\author{Martin H. M\"user$^1$ and Erik Luijten$^2$}

\address{$^1$Institut f\"ur Physik, WA 331,
             Johannes Gutenberg-Universit\"at,
             D-55099 Mainz, Germany \\
         $^2$Institute for Physical Science and Technology,
             University of Maryland, College Park,
             MD 20742-2431, USA
        }
\date{\today}
\maketitle

\begin{abstract}
  The liquid--vapor transition in $^3$He and $^4$He is investigated by means of
  path-integral molecular dynamics and the quantum virial expansion. Both
  methods are applied to the critical isobar and the critical isochore.  While
  previous path-integral simulations have mainly considered the lambda
  transition and superfluid regime in $^4$He, we focus on the vicinity of the
  critical point and obtain good agreement with experimental results for the
  molar volume and the internal energy down to subcritical temperatures.  We
  find that an effective classical potential that properly describes the
  two-particle radial distribution function exhibits a strong temperature
  dependence near the critical temperature.  This contrasts with the behavior
  of essentially classical systems like xenon, where the effective potential is
  independent of temperature. It is conjectured that, owing to this difference
  in behavior between classical and quantum-mechanical systems, the crossover
  behavior observed for helium in the vicinity of the critical point differs
  qualitatively from that of other simple liquids.
\end{abstract}
}

\section{Introduction}
Crossover phenomena have enjoyed a renewed attention in recent years, both from
the theoretical (see, e.g.,
Refs.~\onlinecite{bagnuls85,anisimov92,chicross,caracciolo99,mr3d}) and from
the experimental side (cf.\ 
Refs.~\onlinecite{meier93,anisimov95,melnichenko97}); see also
Ref.~\onlinecite{anisimov00} and references therein.  This concerns in
particular the crossover from mean-field-like to Ising-like critical behavior
upon approach of the critical point.  The accurate numerical determination of
crossover scaling functions for the isothermal compressibility and the
liquid--vapor coexistence curve\cite{chi3d} has motivated the reexamination of
experimental data for $^3$He and Xe.\cite{luijtenmeyer} Xenon, with a very high
molar mass and a relatively high critical temperature, is essentially a
classical system, while the critical point for $^3$He occurs at a temperature
and density where quantum effects are expected to be non-negligible.  This is
also expressed by the de Boer parameter $\Lambda^*$.\cite{deboer48} For
monatomic gases of atomic mass~$m$, that are described by a Lennard-Jones
potential with parameters $\varepsilon$ and $\sigma$, this parameter is defined
as $\Lambda^* = h/\sigma\sqrt{m\varepsilon}$. For $^3$He its value is
$\Lambda^*=3.08$, compared to $\Lambda^*=0.064$ for xenon.\cite{hirschfelder}
Nevertheless, the nature of the critical point itself is the same for both
fluids, because the critical fluctuations dominate over the quantum-mechanical
fluctuations at temperatures sufficiently close to the critical
temperature~$T_c$. Thus, the values of the critical exponents are not affected.
For the crossover region, the situation is less clear-cut.  Since the
correlation length now has a finite (although large) value, nonuniversal
behavior may be expected for different systems and has actually been
observed.\cite{anisimov95} However, according to theory, this nonuniversality
is largely determined by the so-called cut-off parameter (the wavenumber
corresponding to some microscopic characteristic length) which in most {\em
simple\/} fluids has a very similar value.\cite{anisimov95} Furthermore, also
in numerical simulations of Ising-like systems a high degree of universality
has been observed for crossover scaling functions.\cite{dblock} Thus, it came
as quite a surprise that the crossover behavior for $^3$He exhibits a marked
difference from that of xenon.\cite{luijtenmeyer} Owing to the short range of
the interactions, these systems cannot be expected to complete the full
crossover to mean-field-like behavior before leaving the critical region [where
$t \ll 1$, with $t \equiv (T-T_c)/T_c$]. Nevertheless, for $T>T_c$, the
crossover behavior of the isothermal compressibility of xenon turned out to
agree very well with the numerical data for the three-dimensional Ising model
with varying interaction range,\cite{chi3d} whereas the corresponding
experimental data for $^3$He seemed to be described by a {\em qualitatively
different\/} curve. For $t \gtrsim 0.01$, the compressibility appeared
essentially {\em suppressed\/} compared to the crossover scaling function. In
Ref.~\onlinecite{luijtenmeyer}, this difference in behavior was conjectured to
be related to quantum-mechanical effects: The critical compressibility of
$^3$He would be enhanced due to quantum fluctuations, which are temperature
dependent.  Hence, this contribution is expected to decrease appreciably within
the (high-temperature) crossover region, effectively leading to an additional
reduction of the compressibility upon increase of the temperature. Clearly, it
is only the compressibility due to the thermal fluctuations which is described
by the various theoretical expressions for the crossover scaling function.

In order to gauge the quantum-mechanical contribution to the compressibility, a
comprehensive theoretical description of the critical behavior, {\em
including\/} the role of quantum effects, is required; such a description
should encompass the temperature dependence of this contribution in the
vicinity of~$T_c$. Indeed, for a weakly interacting Bose fluid a scaling
function has been calculated describing the crossover from criticality (i.e.,
the lambda point) to ideal Bose-gas behavior.\cite{rasolt84,weichman86} In
particular, it was found that a mapping of the Hamiltonian for the Bose gas
onto a classical spin model yields a Landau-Ginzburg-Wilson Hamiltonian with a
quartic term that has a strongly temperature-dependent
coefficient.\cite{rasolt84} Since this coefficient plays a pivotal role in
crossover scaling functions, a corresponding effect in the vicinity of the
liquid--vapor critical point would definitely affect the nature of the
crossover from Ising-like to mean-field-like critical behavior.  These
considerations have motivated us to examine the behavior of helium close to the
critical point by means of quantum-mechanical numerical methods.  In
particular, it is of interest to see whether the magnitude of quantum effects
indeed changes appreciably over the crossover region, as conjectured in
Ref.~\onlinecite{luijtenmeyer}. In the present study, we pay some attention to
$^3$He, but our main focus is on $^4$He.  Due to the higher mass of $^4$He,
quantum effects are less pronounced, which is only reinforced by the
correspondingly higher critical temperature. This facilitates the numerical
calculations considerably, as will be outlined in Sec.~\ref{sec:pimd}. At the
same time, the quantum effects are still clearly visible in the crossover
region and the isothermal compressibility indeed exhibits a deviation from the
predicted crossover curve, similar to that found for $^3$He.\cite{luijtenmeyer}
For comparison, $^4$He has a de~Boer parameter $\Lambda^*=2.67$, compared to
the above-mentioned value $\Lambda^*=3.08$ for $^3$He.\cite{hirschfelder}

For completeness we remark that $^4$He has been extensively studied by means of
the path-integral Monte Carlo method, especially in the context of the lambda
transition taking place at $2.17$~K, cf.\ Ref.~\onlinecite{ceperley95}.  Also
the high-density region has been explored in this way,\cite{ceperley96} but we
are not aware of numerical studies in the vicinity of the liquid--vapor
critical point.

The outline of this paper is as follows. In Sec.~\ref{sec:methods} we introduce
the methods that we have applied: path-integral molecular dynamics is discussed
in Sec.~\ref{sec:pimd}, followed by the quantum virial expansion in
Sec.~\ref{sec:virial}. Section~\ref{sec:results} contains all our main results,
namely data for the atomic volume of $^4$He and for its kinetic, potential, and
internal energy along the critical isobar, as well as for the kinetic and
potential energy of $^4$He and $^3$He along their respective critical
isochores. Furthermore, results for the effective pair potential of $^3$He in
the vicinity of the critical temperature are presented. Our conclusions are
summarized in Sec.~\ref{sec:conclusion}.

\section{Methods}
\label{sec:methods}

\subsection{Path-integral Molecular Dynamics}
\label{sec:pimd}
Path-integral Monte Carlo (PIMC)\cite{ceperley79,schmidt92} and path-integral
molecular dynamics (PIMD)\cite{tuckerman93} are well-established methods to
calculate thermodynamic properties of many-particle quantum systems. Path
integral techniques exploit the possibility to represent the quantum-mechanical
partition function of point particles as a classical partition function of
closed polymers.\cite{ceperley95} Neighboring beads in the polymer are coupled
via elastic springs with stiffness $k = mM^2/\beta^2\hbar^2$, where $m$ is the
mass of the point particle, the so-called Trotter number $M$ represents the
number of beads in the polymer, and $\beta=1/k_{\rm B}T$.  Interaction between
polymers only takes place between those monomers that have the same index
within the respective polymers they belong to, and the effective temperature in
the isomorphic classical representation is given by $TM$. For more details, we
refer the reader to the original literature and review
articles.\cite{ceperley95,ceperley79,feynman65,marx99}

Depending on the nature of the problem under study, Monte Carlo methods are
preferable to molecular dynamics or vice versa.  PIMC has so far been the
method of choice for finite-temperature simulations of condensed helium, in
particular when exchange effects played an important role. Far away from the
superfluid regime, however, exchange effects can be neglected,\cite{ceperley95}
so that PIMC is not necessarily advantageous to PIMD for the study of the
liquid--vapor transition. For example, in the gas phase, ballistic trajectories
can be realized in PIMD, which allows rapid changes in the configuration.  In
conventional Monte Carlo methods, motions are constrained to be diffusive and
large correlation times may be expected.  Also, in isobaric simulations, volume
moves are done at no extra cost in a molecular dynamics simulation, while Monte
Carlo methods require the evaluation of the total energy of the system.

Using an appropriate representation of the internal coordinates of the chain
molecules, it is possible to avoid inefficient sampling\cite{tuckerman93};
e.g., in the regular representation, where each bead in a chain has the same
dynamical inertia, the time-step discretization has to be chosen proportional
to $M^{-2}$.  In this study, we have used a representation of the chain
molecules in terms of the center-of-mass coordinate and the eigencoordinates of
the free particles, defined in Eq.~(\ref{eq:eigen_coord}) below, which makes it
possible to work with time-steps independent of $M$. While the ``dynamical''
center-of-mass coordinate is chosen to be identical with the real mass~$m$,
different masses~$m_q$ are attributed to each eigenmode~$q$.  An efficient
choice is $m_q = (\tilde{k} + k_q) m / \tilde{k}$, where $\tilde{k}$ is an
adjustable parameter and $k_q$ is the stiffness associated with the
eigenmode~$q$. $\tilde{k}$ is conveniently chosen such that in the condensed
phase all modes move on approximately the same time scale.

The main disadvantage of PIMD is that finite-time step errors accumulate for
certain quantities when the Trotter number~$M$ becomes large. While the average
potential energy and the virial estimator for the kinetic energy\cite{herman82}
do not suffer from this effect, the so-called primitive estimator for the
kinetic energy, $K_{\rm prim}$, does.  An estimator is a function whose {\em
average\/} value corresponds to the expectation value of a property of
interest.  For an {\em individual\/} configuration, however, the association of
the actual value of the estimator with the actual value of the property is
meaningless.  $K_{\rm prim}$ is given by\cite{herman82}
\begin{equation}
K_{\rm prim} = {3 \over 2} N k_{\rm B} T M^2 - {k \over 2} \sum_{i=1}^N
\sum_{\tau=1}^M \left( {\bf r}_{i\tau} - {\bf r}_{i\tau+1}\right)^2 \;,
\label{eq:primit}
\end{equation}
where ${\bf r}_{i\tau}$ represents the position of the $\tau$th monomer in
polymer~$i$, and $k$ is the stiffness introduced at the beginning of this
section.  In addition to the above-mentioned systematic error, $K_{\rm prim}$
has been shown to have large statistical errors for large Trotter numbers~$M$,
even in PIMC simulations.\cite{herman82} However, a simple trick remedies both
of these shortcomings.\cite{schoffel01} If ${3 \over 2} N k_{\rm B} T M^2$ is
replaced by the actual ``dynamic'' kinetic energy, a new estimator can be
defined,
\begin{equation}
\tilde{K}_{\rm prim} =  {1\over 2}\sum_{i=1}^N  \sum_{q=1}^M
\left(m_q {\bf v}^{\prime 2}_{iq} - k_q {\bf r}^{\prime 2}_{iq} \right) \;,
\label{eq:prim_new}
\end{equation}
where ${\bf v}'_{iq}$ denotes the velocity associated with the eigencoordinate
\begin{equation}
{\bf r}'_{iq} = {1\over \sqrt{M}} \sum_{\tau=1}^M {\bf r}_{i\tau}
e^{2\pi i q \tau /M} \;.
\label{eq:eigen_coord}
\end{equation}
It turns out that the new estimator $\tilde{K}_{\rm prim}$ is superior to the
virial estimator.  In the solid phase, e.g., the virial estimator misses the
center-of-mass motion of the simulation box, leading to a systematic difference
between the primitive and the virial estimator of $3k_B T/2$ for the whole
system.  While this effect is negligible for large particle numbers, the same
shortcoming has more serious implications in the gas phase: The virial
estimator systematically underestimates the kinetic energies because it does
not include the kinetic energy of the center-of-mass motion of independent
clusters.  In the condensed phase, the statistical errors of both estimators
are similar.\cite{schoffel01} Unfortunately, $\tilde{K}_{\rm prim}$ cannot be
used in Monte Carlo~(MC) simulations, because MC is not based on deterministic
dynamics but solely on stochastic dynamics.  It is worth mentioning that the
average of both estimators tends to the exact thermal expectation value of the
kinetic energy with increasing Trotter number~$M$.  For methods to perform PIMD
simulations at constant pressure, we refer to
Refs.~\onlinecite{schoffel01,martyna99}.

All simulations, as well as the virial expansions, are based on the Aziz HFD-B
potential,\cite{aziz87} which is considered one of the best known interatomic
model potentials.\cite{ceperley95} It consists of a Hartree--Fock (exponential)
short-range repulsive term and algebraic long-range attractive terms ($1/r^6$,
$1/r^8$, $1/r^{10}$).  The cut-off radius that we used in our simulations was
$r_{\rm c} = 10$~\AA.  The particle number in all simulations was $N = 500$ and
the Trotter number varied between $M = 1$ for classical simulations and $M =
64$ for the quantum-mechanical $^4$He simulations at the lowest temperatures,
keeping $TM > 200$~K. The usual corrections of the order of $1/M^2$ were
applied to the final data.\cite{suzuki87} For the $^3$He simulations, $TM >
350$~K was used with a maximum value of $M = 128$. We have used a cubic
simulation box with periodic boundary conditions; the length of each simulation
amounted to at least 50\,000 time steps.

Note that our simulations mostly address the identification of quantum effects
in the first derivatives of the thermodynamic potential (internal energy,
volume, etc.), while an estimation of second derivatives such as the
compressibility (discussed in Ref.~\onlinecite{luijtenmeyer}) is not attempted:
This would require much larger system sizes and is not feasible with our
computer resources.

\subsection{Virial Expansion}
\label{sec:virial}

An alternative method for calculating thermodynamic properties of quantum gases
is by means of the quantum virial expansion.  In the original approach (see
Ref.~\onlinecite{hirschfelder}, Chapter~6 and references therein), a series
expansion in $\hbar$ is obtained for every virial coefficient. Here, a more
efficient technique is used,\cite{storer68,thirumalai83} which is briefly
outlined below.

The first correction to the internal energy of an ideal gas arises from the
pair interaction $u_{12}(\beta,V)$ of (quantum-mechanical) particles.  A pair
of particles is described by a center of mass mode, which can be treated
classically in the second-order virial expansion, and the relative coordinate
${\bf r}$, which is confined to a (spherical) volume $V$.  Throughout the
derivation, finite-volume corrections are ignored.  For $N^{2}/2$ particle
pairs, $U(N,V,\beta)$ becomes\cite{ceperley95}
\begin{equation}
U(N,V,\beta) = {3 N \over 2 \beta} + {N^2 \over 2}
u_{12} (\beta,V) + {\cal O}\left\{N\left(\frac{N}{V}\right)^2\right\} \;,
\end{equation}
where $u_{12}(\beta,V)$ is calculated according to
\begin{eqnarray}
u_{12} (\beta,V) &=&
- { \partial \over \partial \beta }
\log \int_{V} d^3 r \, \langle {\bf r} \mid
e^{ -\beta ( \hat{t}_{\rm rel} + \hat{v}_{\rm 12} ) } \mid {\bf r} \rangle
\nonumber \\
& & + { \partial \over \partial \beta }
\log \int_{V} d^3 r \, \langle {\bf r} \mid
e^{ -\beta  \hat{t}_{\rm rel}  }  \mid {\bf r} \rangle \;.
\label{eq:rel_energ}
\end{eqnarray}
where $\hat{t}_{\rm rel}$ denotes the operator for the kinetic energy
associated with the {\em relative\/} motion of two particles and $v_{12}$ is
their potential energy.  The two integrands on the right-hand side of
Eq.~(\ref{eq:rel_energ}) are the diagonal elements of the density matrix
$\rho({\bf r}, {\bf r}', \beta)$,
\begin{equation}
\rho({\bf r}, {\bf r}', \beta) =
\langle {\bf r} \mid \exp[-\beta (\hat{t}_{\rm rel} + \hat{v}_{12})]
\mid {\bf r}'\rangle \;,
\end{equation}
and of its noninteracting counterpart.

Up to a normalization factor, which is irrelevant for the calculation of
$u_{12}(\beta,V)$ in Eq.~(\ref{eq:rel_energ}), the radial distribution function
$g(r)$ is given by the diagonal elements of $\rho({\bf r}, {\bf r}', \beta)$.
Taking into account that $\int_V d^3 r\, g(r) \to V$ in the thermodynamic
limit, it is possible to rewrite Eq.~(\ref{eq:rel_energ}) in the more familiar
form
\begin{equation}
u_{12}(\beta, V) = {2\over V} {\partial B_2(\beta) \over \partial \beta},
\end{equation}
where, like for classical systems, the second virial coefficient $B_2(\beta)$
can be expressed in terms of $g(r)$:
\begin{equation}
B_2(\beta) = -2\pi \int_0^\infty dr \, r^2 \,
[g_{12}(r)-g_{0}(r)] \;.
\label{eq:vir_coef}
\end{equation}
Here $g_{12}(r)$ and $g_{0}(r)$ denote the radial distribution function in the
interacting and non-interacting case, respectively.

The diagonal elements $\rho({\bf r}, {\bf r}, \beta)$ can be calculated by
exploiting the semi-group property of the density operator
\begin{equation}
\rho({\bf r}, {\bf r}', 2\beta) =
\int d^3 r'' \rho({\bf r}, {\bf r}'', \beta)
             \rho({\bf r}'', {\bf r}', \beta) \;.
\label{eq:square}
\end{equation}
Thus we can obtain the low-temperature density matrix at temperature $T$ by
squaring the density matrix at temperature $2T$. For $n$ iterations, the
starting temperature has to be chosen as $2^n T$.  For the highest temperature
of the iteration process, it is possible to use the so-called primitive
decomposition for $\rho({\bf r}, {\bf r}', \beta/M)$, which underlies the
path-integral simulations presented in this paper as well as most other
path-integral simulations.  One of the advantages of the squaring procedure
over path-integral simulations is that the required numerical effort only
scales logarithmically with inverse temperature.  Hence, it is easy to minimize
discretization errors (for two-particle systems) using squaring techniques at
low temperatures.  At a given (high) temperature, the systematic error is
proportional to $1/M^2$, just like in path-integral simulations.\cite{suzuki87}

For spherically symmetrical potentials, Eq.~(\ref{eq:square}) can be reduced to
a sum of one-dimensional integrations by decomposing the density matrix into
contributions belonging to different angular
momenta.\cite{storer68,thirumalai83} In practice, the squaring is done in terms
of simple matrix multiplication by discretizing the variable $r$.  Of course, a
cut-off $r_c$ has to be introduced at a reasonably large value of $r$.  This
induces artificial behavior at the boundary not found in an infinitely large
system, namely that $g_0(r)$ tends to zero as $r$ approaches $r_c$.  Therefore,
the integration in Eq.~(\ref{eq:vir_coef}) has to be confined to the region
where boundary effects are negligible.  Alternatively, one may normalize the
integrand in Eq.~(\ref{eq:vir_coef}) by $g_0(r)$, which results in a fast
convergence of $B_2$ with $r_c$.

\begin{figure}
\leavevmode\centering
\epsfxsize=80mm
\epsffile{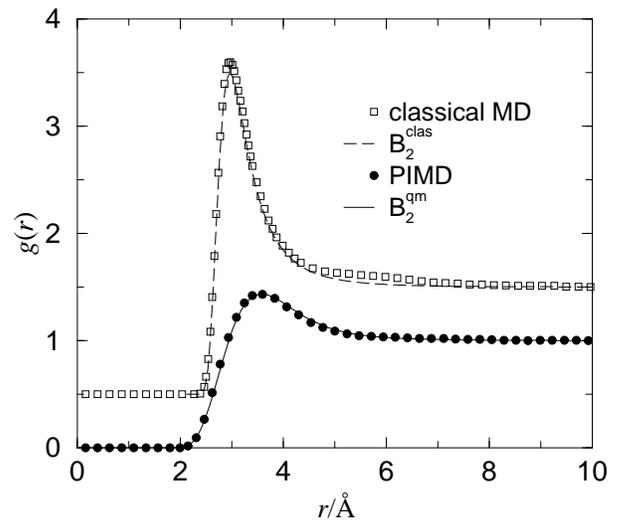}

\caption{Radial distribution function $g(r)$ for $^4$He at temperature $T =
10$~K and pressure $P = 0.22746$~MPa, as obtained from both classical and
quantum-mechanical calculations. The points result from simulations, while
lines indicate second-order virial expansions. For clarity, the classical
curves have been raised by 0.5.}
\label{fig:g_r10K}
\end{figure}

Quantum effects in the calculation of $B_2(\beta)$ will become important when
the thermal wavelength $\lambda(\beta) = h/\sqrt{2\pi mk_{\rm B}T}$ of the free
particle is in the order of or larger than the distance at which the
interatomic potential is minimum.  We want to illustrate this in the case of
$^4$He at a temperature $T = 10$~K, where $\lambda \approx 2.8$~\AA.  In
Fig.~\ref{fig:g_r10K}, the two-particle radial distribution function $g(r)$ is
shown for a pair of ``classical'' helium atoms and a pair of $^4$He atoms.  The
maximum in $g(r)$ for the quantum-mechanical calculation is shifted by about
$0.59$~\AA\ with respect to the classical equilibrium distance and the height
of the maximum (relative to $g_0 = 1$) is decreased by a factor of about 5.8.
Thus, an effective classical potential $V_{\rm eff}(r)$ that would result in a
similar $g(r)$ of $^4$He at $T = 10$~K as the quantum-mechanical calculation
would have a strongly reduced binding energy with respect to the original Aziz
potential and an equilibrium distance shifted by $0.59$~\AA.  Note that also
the curvature of the effective potential would be different from the original
potential.

In Fig.~\ref{fig:g_r10K}, PIMD simulation results, taken at the same
temperature and the critical pressure $P_{c,4} = 0.22746$~MPa\footnote{The
subscript `4' will generally be used to indicate quantities pertaining to
$^4$He, whereas a subscript `3' refers to $^3$He.}, are included as well.  The
agreement between the virial expansion and the simulation is very good. For the
classical system, small differences in $g(r)$ can be seen between both methods
that can be attributed to three-body effects, which are neglected in the
second-order virial expansion.

\section{Results}
\label{sec:results}

\subsection{Critical isobar of $^4$He}

\begin{figure}
\leavevmode\centering
\epsfxsize=80mm
\epsffile{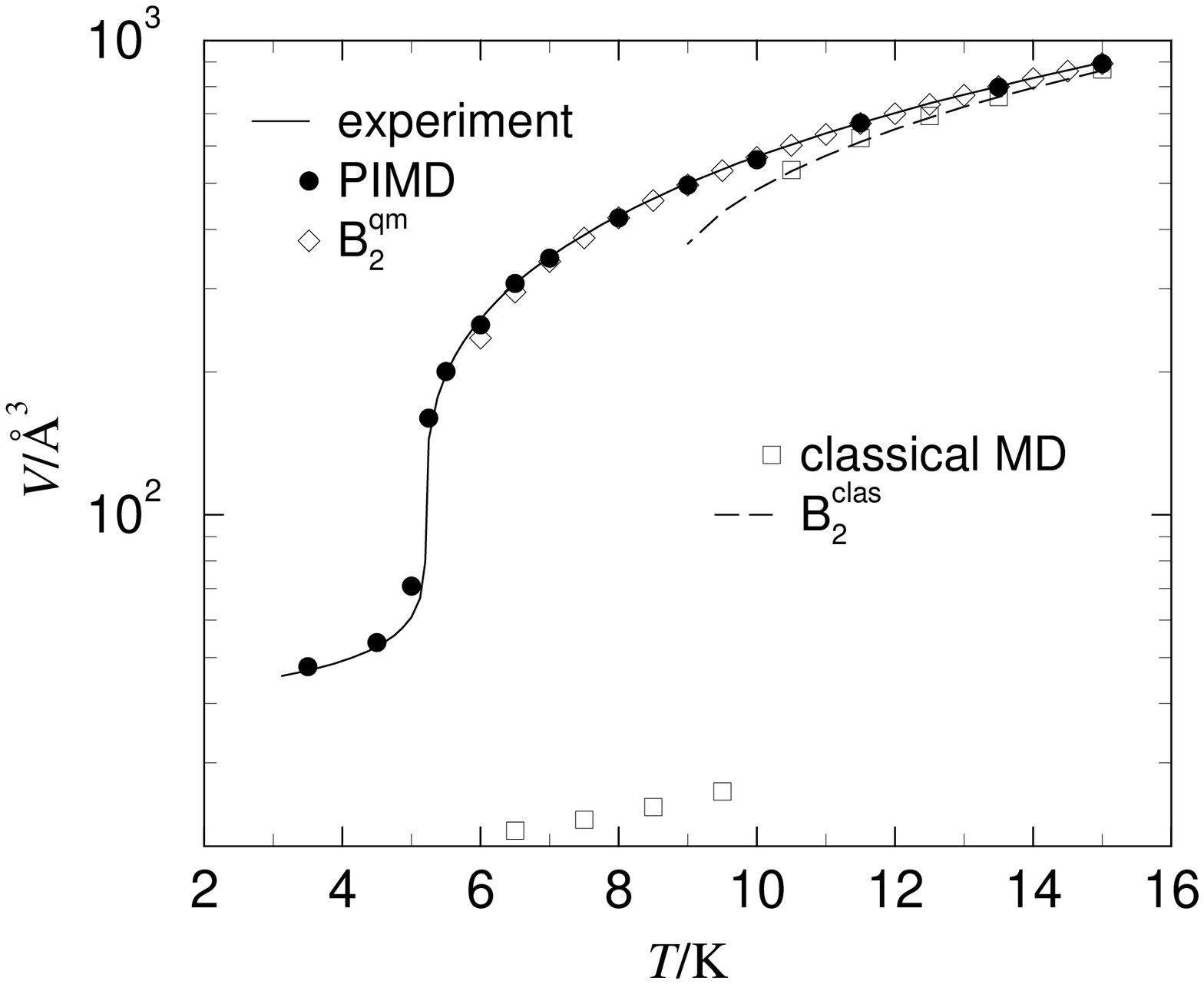}

(a)

\leavevmode\centering
\epsfxsize=80mm
\epsffile{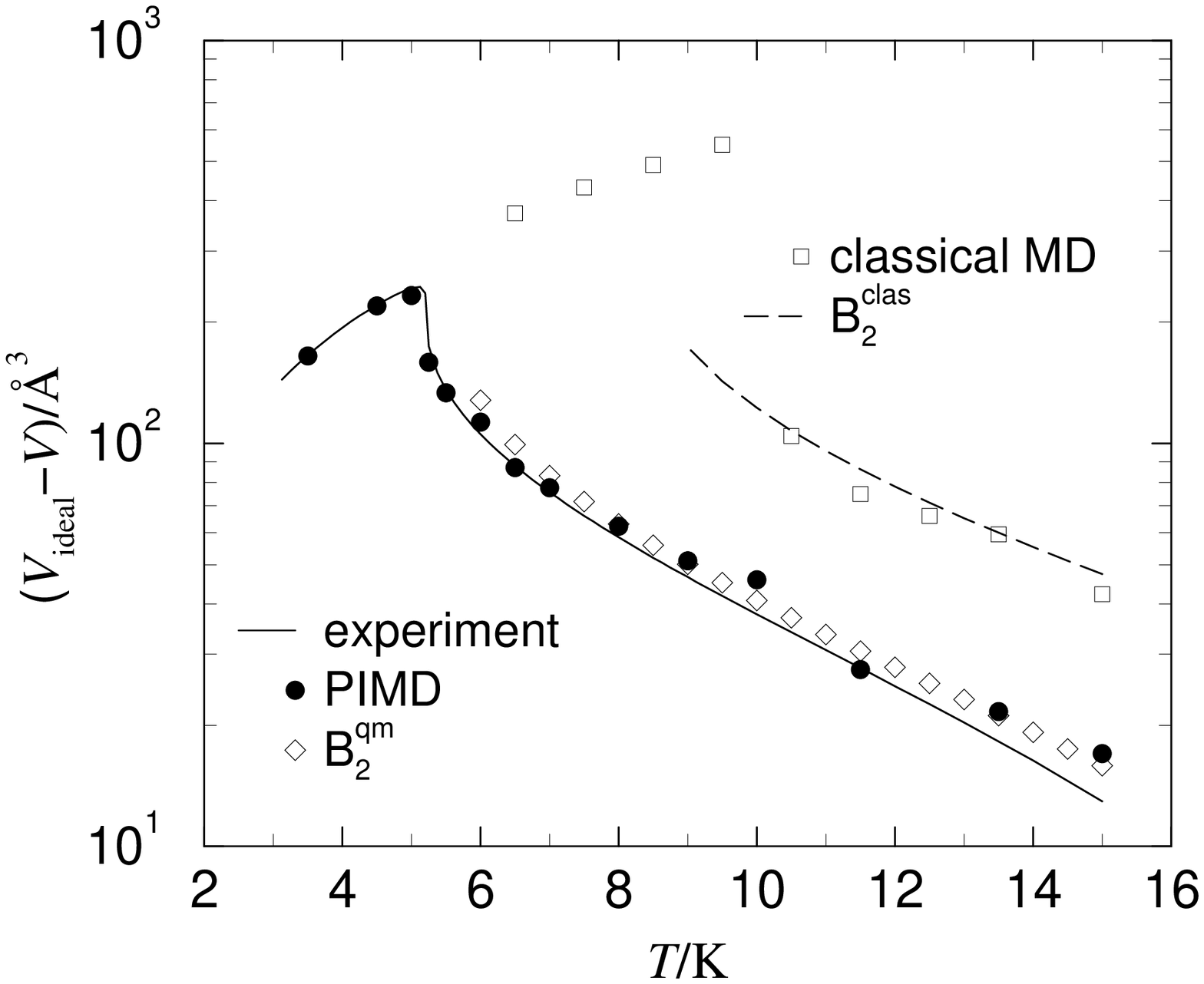}

(b)

\caption{(a) Volume~$V$ per atom for $^4$He (experiments correspond to the
solid line, PIMD data to the circles, and the second-order quantum-mechanical
virial expansion to the diamonds) and for ``classical helium'' (classical MD
data correspond to the open squares and the classical virial expansion to the
dashed line) as a function of temperature at the critical isobar of $^4$He
($P_{c,4} = 0.22746$~MPa).  (b) The same symbols as in a), but now the
difference between the ideal gas volume $V_{\rm ideal} = k_{\rm B} T / P$ and
the actual volume per atom is shown as a function of temperature.}
\label{fig:volume}
\end{figure}

As a first test of the PIMD simulation and the virial expansion, we have
calculated the atomic volume as a function of temperature on the critical
isobar, $P_{c,4} = 0.22746$~MPa.\cite{nist} While for $^3$He the diameter
$\rho_d = (\rho_{\rm liq} + \rho_{\rm vapor})/2$ has a slope that is almost
zero,\cite{pittman79} for $^4$He it has a clearly positive slope.  Thus, while
the isobaric thermal expansion coefficient $V^{-1}(\partial V/\partial T)_P$
only has a finite peak as a function of temperature for $P>P_c$, it diverges at
the liquid--vapor transition temperature for $P \leq P_c$.  The results for
classical and quantum-mechanical calculations are presented in
Fig.~\ref{fig:volume} and comparison is made to a phenomenological wide-range
equation of state based on experimental data.\cite{nist,mccarty90} As can be
seen, the overall agreement between the experimental results and the
quantum-mechanical simulations is very good.  This demonstrates both the
quality of the Aziz potential and the fact that the (neglected) three-body
effects play a remarkably small role here.  Only near the critical temperature,
$T_{c,4} = 5.1953$~K,\cite{nist} small systematic deviations occur, which
however may be tentatively attributed to finite-size effects in the
simulations.  In addition, the performance of the phenomenological equation of
state itself might deteriorate in the critical region.  The second-order
quantum virial expansion, included in the same figure, exhibits deviations from
the experimental curve below a temperature of approximately 7~K.

For comparison, data for ``classical helium'' are included in
Fig.~\ref{fig:volume} as well. The absence of quantum fluctuations leads one to
expect an increase in both the critical temperature and the critical pressure.
Thus, the classical system should undergo a first-order phase transition at the
critical pressure~$P_{c,4}$, as is indeed borne out by the numerical data.  The
corresponding transition temperature~$T_1$ could be located at approximately
$10$~K; here both the fluid and the gas phase were stable for the duration of
the simulation. Hysteresis effects and, possibly, slowing down owing to the
vicinity of the critical point, may have affected the accuracy of this
estimate. Since neither the pressure~$P_{c,4}$ nor the corresponding transition
temperature play a particular role for the classical system, no further
attempts were made to improve the estimate of~$T_1$. We note only that the {\em
critical\/} temperature $T_c^{\rm class}>T_{c,4}$ of the classical system will
be even higher than~$T_1$.  The magnitude of the shift of the transition
temperature is a clear indication of the importance of quantum-mechanical
effects in the vicinity of the critical point. Indeed, from the
principle of corresponding states~\cite{hirschfelder} one finds (approximating
the interaction potential by a Lennard-Jones potential with $\sigma=2.56$~\AA\
and $\varepsilon/k_{\rm B}=10.22$~K, Ref.~\onlinecite{hirschfelder}) $T_c^{\rm
class}=13$~K and $P_c^{\rm class}=1.1$~MPa.

\begin{figure}
\leavevmode\centering
\epsfxsize=80mm
\epsffile{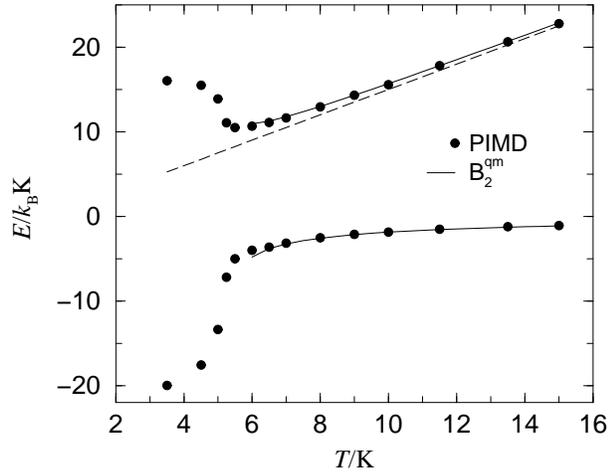}
\caption{Kinetic energy (positive values) and potential energy (negative
values) for $^4$He as a function of temperature $T$, at a pressure $P =
0.22746$~MPa. Points result from simulations, while the solid curves result
from the second-order virial expansion.  The dashed line indicates the
classical kinetic energy.}
\label{fig:eneqm_bar}
\end{figure}

An interesting effect can be observed in the kinetic energy~$T_{\rm kin}$ for
$^4$He, shown in Fig.~\ref{fig:eneqm_bar}. Upon lowering the temperature,
$T_{\rm kin}$ suddenly rises in the vicinity of the critical temperature.  This
rise is purely related to the increase in the density of the system. It should
be noted that the behavior of $T_{\rm kin}$ close to $T_c$ is likely to be
affected by finite-size effects, as these effects generally shift the top of
the coexistence curve toward temperatures {\em above\/} the true critical
temperature.  The break-down of the second-order virial expansion due to this
increase in density occurs already slightly above $T_{c,4}$, namely around
5.5~K.  The potential energy~$V_{\rm pot}$, depicted in the same figure,
increases with temperature, as usual. We note, however, that this tendency of
decreasing kinetic energy and increasing potential energy upon increase of the
temperature near~$T_c$ is not a necessity: A quantum-mechanical model for
molecular ordering of rotors on a surface rather showed the opposite trend,
where $\langle T_{\rm kin}\rangle$ increased and $\langle V_{\rm pot}\rangle$
decreased near~$T_c$ upon increasing temperature.\cite{hetenyi99}

\begin{figure}
\leavevmode\centering
\epsfxsize=80mm
\epsffile{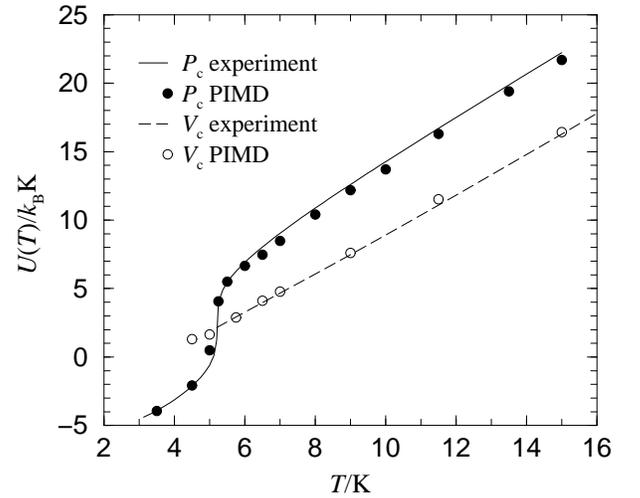}
\caption{A comparison of experimental and numerical results for the internal
energy of $^4$He as a function of temperature, both along the critical isobar
(solid curve and closed circles, respectively) and along the critical isochore
(dashed line and open circles, respectively). In both cases, the agreement is
good.}
\label{fig:ene_net}
\end{figure}

Figure~\ref{fig:ene_net} shows the {\em internal\/} energy along the critical
isobar, both as obtained from experiment\cite{nist,mccarty90} and as calculated
by means of PIMD\@. Although the experimental data lie systematically above the
numerical ones for $T>6$~K, the overall agreement is certainly appreciable.

Finally, we mention that the average {\em classical\/} potential energy
$\langle V_{\rm pot}\rangle$, which is not shown in Fig.~\ref{fig:eneqm_bar},
exhibits a clear jump, as expected for a first-order transition.

\subsection{Critical isochore of $^4$He}

\begin{figure}
\leavevmode\centering
\epsfxsize=80mm
\epsffile{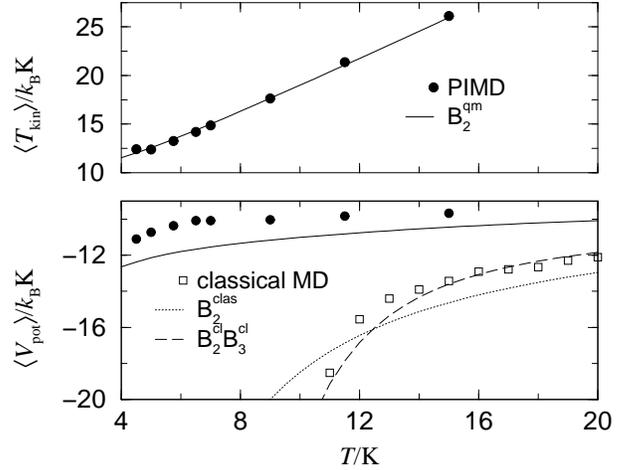}
\caption{The kinetic (positive values) and potential (negative values) energy
per atom for $^4$He as a function of temperature~$T$, along its critical
isochore.  Points result from simulations (closed circles for the
quantum-mechanical calculations and open squares for the classical ones), while
the solid curves result from second-order virial expansions. The dotted and the
dashed curves indicate second- and third-order virial expansions, respectively,
for the classical potential energy.}
\label{fig:ene_cho}
\end{figure}

We now turn to the critical isochore, $\rho_{c,4} = 0.017399$
mol/cm$^3$.\cite{nist} Figure~\ref{fig:ene_cho} shows both the kinetic and the
potential energy per atom, along with the results of the quantum virial
expansion. As can be seen, the agreement is remarkably good for the kinetic
energy, even in the critical region.  For the potential energy, on the other
hand, the agreement is not so good, even at relatively high temperatures. In
the same figure, we have also included the potential energy for ``classical
helium'' as obtained from a second-order and the third-order virial expansion:
Here the agreement is much better.  However, a remark on the density is in
order here. As mentioned before, quantum fluctuations generally lead to a lower
critical temperature and hence, at the same pressure, to a higher density.
However, the {\em critical\/} pressure decreases, and the net effect is that
fluids in which quantum-mechanical effects are nonnegligible have a {\em
lower\/} critical density than would be expected from the principle of
corresponding states.\cite{hirschfelder} This is also nicely illustrated by the
critical properties of $^3$He, which is basically described by the same pair
potential as $^4$He, but has an even lower critical temperature, pressure, and
density, entirely due to its lower mass and consequentially larger de Boer
parameter.\cite{deboer48} Indeed, a quantum-mechanical version of the principle
of corresponding states can be formulated, in which all deviations from
classical behavior are parametrized by this parameter.\cite{hirschfelder} From
this slight digression, we conclude that the classical data, as shown in the
figure, pertain to an isochore that is (for the {\em classical\/} system) a
subcritical one, which might (in addition to the higher order of the virial
expansion) explain the quite reasonable agreement. One remaining point, then,
is that this isochore must cross the vapor branch of the coexistence curve at
an unknown temperature.

Just like along the critical isobar, the PIMD results for the internal energy
along the critical isochore (Fig.~\ref{fig:ene_net}) exhibit good agreement
with the experimental data.

\subsection{Critical isochore of $^3$He}

\begin{figure}
\leavevmode\centering
\epsfxsize=80mm
\epsffile{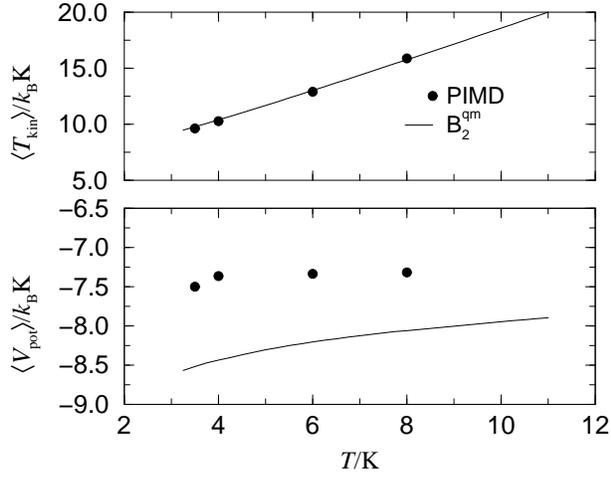}
\caption{The kinetic (positive values) and potential (negative values) energy
per atom for $^3$He as a function of temperature, along its critical isochore.
The closed circles were obtained by means of simulations and the solid curves
represent second-order virial expansions.}
\label{fig:ene_cho_he3}
\end{figure}

Finally, we consider $^3$He on its critical isochore, $\rho_{c,3} =
0.01374$~mol/cm$^3$.\cite{pittman79} Figure~\ref{fig:ene_cho_he3} shows both
the kinetic and the potential energy per atom as obtained from PIMD simulations
and from the second-order quantum virial expansion. As for $^4$He, the
agreement between both types of calculations is very good for the kinetic
energy and rather poor for the potential energy.  Also the overall behavior of
both energies is similar to that found for $^4$He, except that in
Fig.~\ref{fig:ene_cho_he3} one cannot observe the formation of a ``plateau'' in
$\langle T_{\rm kin} \rangle$ at lower temperatures. This is presumably due to
the fact that the simulations for $^3$He do not quite reach the critical
temperature, $T_{c,3} = 3.317$~K (see Ref.~\onlinecite{pittman79}; the value
was converted to the $T_{90}$ temperature scale here), where the flattening of
the curve is expected to set in.

\begin{figure}
\leavevmode\centering
\epsfxsize=80mm
\epsffile{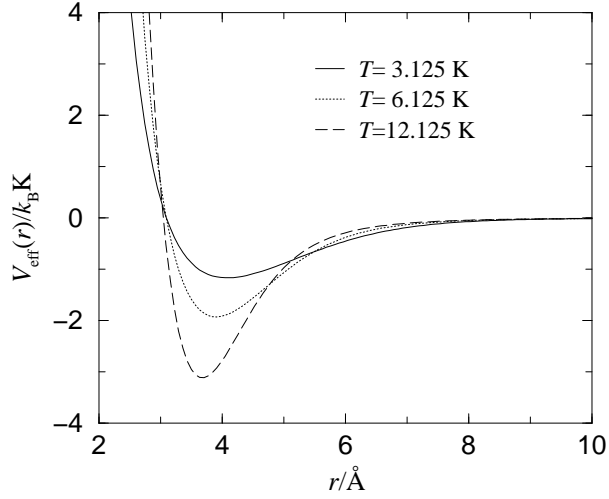}
\caption{Effective potential for $^3$He as obtained from the quantum virial
expansion, for three different temperatures. One observes the deepening and
narrowing of the potential well when the temperature is increased from $T
\simeq T_c$ to $T \simeq 4T_{c,3}$. The Aziz HFD-B
potential~\protect\cite{aziz87} takes its minimum at $r=2.963$~\AA.}
\label{fig:vpot_eff}
\end{figure}

In principle, one might use the kinetic energy to define an effective,
nonlinear temperature scale in which the role of the quantum fluctuations has
been taken into account. By studying the crossover scaling function for, e.g.,
the compressibility on such a redefined temperature scale, one could examine
the role of quantum effects in the deviations observed in
Ref.~\onlinecite{luijtenmeyer}.  However, the definition of such a temperature
scale requires a very accurate knowledge of the kinetic energy, in particular
for temperatures very close to~$T_{\rm c}$, since crossover scaling functions
are studied on a logarithmic scale in the reduced temperature~$t$.
Unfortunately, the numerical accuracy of our PIMD data did not warrant a
meaningful, direct reexamination of the crossover scaling functions.  As an
alternative, we provide here qualitative evidence justifying the conjecture of
Ref.~\onlinecite{luijtenmeyer} that the influence of quantum effects changes
appreciably within the crossover regime.  To this end, we have considered the
effective potential $V_{\rm eff}$ as defined by
\begin{equation}
e^{-\beta V_{\rm eff}(r)} \equiv g(r) \;,
\label{eq:veff_pot}
\end{equation}
where $g(r)$ is the two-particle correlation function. The latter quantity, in
turn, can be obtained from the virial expansion. Both $V_{\rm eff}$ and $g(r)$
are temperature-dependent quantities.  The resulting effective potential is
shown in Fig.~\ref{fig:vpot_eff} for three different temperatures, namely
$T=3.125$~K (just below $T_{\rm c}$), $T=6.125$~K (roughly twice $T_{\rm c}$),
and $T=12.125$~K (somewhat above the highest temperature studied in
Ref.~\onlinecite{luijtenmeyer}).  One observes that the effective potential
changes rather dramatically under this temperature variation: The depth of the
potential well decreases by roughly a factor of~3 when the temperature is
reduced from 12.125 to 3.125~K, and its width increases accordingly. This is in
concordance with the observed enhancement of the compressibility over this
temperature region.

\section{Discussion and Conclusion}
\label{sec:conclusion}
In this paper, we have presented the results of a numerical study of $^4$He and
$^3$He in the vicinity of their respective liquid--vapor critical points. Both
path-integral molecular dynamics and quantum virial expansions have been
applied.

For $^4$He, we have shown that the Aziz potential\cite{aziz87} employed in the
PIMD calculations yields very good agreement with experimental
results\cite{nist,mccarty90} for the atomic volume at the critical isobar over
a wide temperature range, including temperatures that lie considerably below
the critical temperature. Equally good agreement is found for the internal
energy. The second-order quantum virial expansion shows good agreement for the
atomic volume down to roughly two Kelvin above the critical temperature.  The
simulations also demonstrate the importance of quantum effects in $^4$He in the
vicinity of its critical point: A classical system with the same potential has
a critical temperature that is more than twice as high.

In addition, we have studied the kinetic and potential energy of $^4$He and
$^3$He along their critical isochores. For the kinetic energy, there is good
agreement between the quantum virial expansion and the simulation results in
both cases, down to quite low temperatures, but for the potential energy the
agreement is not so good. Higher-order terms in the virial expansion would
probably alleviate the discrepancy, but are tedious to calculate. On the other
hand, simulation results for the isochoric internal energy of $^4$He agree well
with experimental results.

One of our original goals was the definition of an effective, nonlinear
temperature scale based on the kinetic energy. Such a temperature scale might
prove useful for the reexamination of the crossover scaling functions of
helium, which were experimentally observed to differ qualitatively from their
counterparts for other simple liquids; these deviations have been conjectured
to originate from quantum-mechanical effects.\cite{luijtenmeyer} Indeed, since
(part of) these effects would now be incorporated in the new temperature scale,
the crossover scaling functions should exhibit a better agreement with the
(classical) prediction if the conjecture were correct.  However, it is very
difficult to reach the required numerical accuracy in the kinetic energy; in
addition, very large particle numbers are needed in the immediate vicinity of
the critical point, in order to circumvent finite-size effects. Thus, we have
opted for a different strategy, namely the calculation of the effective pair
potential for $^3$He by means of the quantum virial expansion evaluated on the
critical isochore. We find that this pair potential exhibits a strong variation
with temperature in the crossover region. This is in good accord with the
conjecture of Ref.~\onlinecite{luijtenmeyer}: Namely, not only are there strong
quantum effects in $^3$He close to its liquid--vapor critical point, as was
already evident from the values of the critical amplitudes (cf., e.g.,
Ref.~\onlinecite{pittman79}), but these quantum effects also {\em vary\/}
considerably within the crossover region. Such a variation might then explain
the observed apparent depression (relative to the predicted crossover scaling
function) of the compressibility above $T_c$ when the temperature is increased.

\acknowledgments
E.L. acknowledges stimulating comments by Michael E. Fisher, as well as support
from the National Science Foundation (through Grant No.\ CHE 99-81772 to M.~E.
Fisher).  M.H.M. thanks Kurt Binder for useful discussions and acknowledges
support from the BMBF through Grant 03N6015 and from the
Materialwissenschaftliche Forschungszentrum.


\end{document}